# Solving petrological problems through machine learning: the study case of tectonic discrimination using geochemical and isotopic data

Maurizio Petrelli[*] and Diego Perugini

*Department of Physics and Geology, University of Perugia, Piazza Università, 06100, Perugia (Italy)*

[*]Corresponding Author:

E-mail: maurizio.petrelli@unipg.it

Tel.: +39 075 5852607

Fax: +39 075 5852630

**Abstract**

Machine learning methods are evaluated to study the intriguing and debated topic of discrimination among different tectonic environments using geochemical and isotopic data. Volcanic rocks characterized by a whole geochemical signature of major elements ($SiO_2$, $TiO_2$, $Al_2O_3$, $Fe_2O_{3T}$, $CaO$, $MgO$, $Na_2O$, $K_2O$), selected trace elements (Sr, Ba, Rb, Zr, Nb, La, Ce, Nd, Hf, Sm, Gd, Y, Yb, Lu, Ta, Th) and isotopes ($^{206}Pb/^{204}Pb$, $^{207}Pb/^{204}Pb$, $^{208}Pb/^{204}Pb$, $^{87}Sr/^{86}Sr$ and $^{143}Nd/^{144}Nd$) have been extracted from open-access and comprehensive petrological databases (i.e. PetDB and GEOROC). The obtained dataset has been analyzed using support vector machines, a set of supervised machine learning methods, which are considered particularly powerful in classification problems.

Results from the application of the machine learning methods show that the combined use of major, trace elements and isotopes allow associating the geochemical composition of rocks to the relative tectonic setting with high classification scores (93%, on average). The lowest scores are recorded from volcanic rocks deriving from back-arc basins (65%). All the other tectonic settings display higher classification scores, with oceanic islands reaching values up to 99%.

Results of this study could have a significant impact in other petrological studies potentially opening new perspectives for petrologists and geochemists. Other examples of applications include the development of more robust geo-thermometers and geo-barometers and the recognition of volcanic sources for tephra layers in tephro-chronological studies.

**Introduction**

Machine learning (ML) entails the use of algorithms and techniques to detect patterns from large datasets and to exploit the uncovered patterns to predict future trends, classify, or perform other kind of strategic decisions (Murphy 2012). The field of ML has progressed dramatically over the past two decades, developing from a "numerical curiosity" to a practical technology with widespread scientific and commercial use (Jordan and Mitchell 2015). For example, ML is now successfully utilized in several fields like computer vision, speech recognition, natural language processing and robot control (Jordan and Mitchell 2015).

In principle, each complex problem characterized by a large enough number of input samples is well suited for ML applications (Jordan and Mitchell 2015). It is notable that the application of ML techniques has been quite extensively tested in the Earth Sciences (Huang et al. 2002; Petrelli et al. 2003; Masotti et al. 2006; Cannata et al. 2011; Zuo and Carranza 2011; Abedi et al. 2012; Goldstein and Coco 2014) but, surprisingly, their use is still virtually unexplored with regards to the solution of petrological problems. One intriguing and debated petrological application, potentially well-suited for the investigation by ML, is the tectonic discrimination of magmas using geochemical data (Li et al. 2015).

Trace element discrimination diagrams were introduced in the 1970s as a method for identifying the tectonic setting of basalts and other volcanic rocks (Pearce and Cann 1973). At that time, classification diagrams utilized only a few elements plotted as binary or triangular diagrams (Pearce and Cann 1973; Pearce 1976; Pearce and Norry 1979; Wood 1980; Shervais 1982; Meschede 1986; Grimes et al. 2015). This approach is still widely used; to date the work by Pierce and Cann (1973) received about 2082 citations (more than 400 only in the last 5 years; source: Scopus,

August 2016) testifying for the popularity of this approach in the petrological community (Li et al. 2015).

In 2006, Snow (2006) demonstrated that the success of these diagrams is mainly hindered by their limited dimensionality due to visualization requirements. Snow (2006) proposed alternative probabilistic methods and reported single analysis classification success rates for volcanic rocks from island arcs, ocean islands and mid-ocean ridges environments of about 83%, 75% and 76%, respectively.

In addition, Vermeesch (2006a) firstly proposed the application of two dimensional linear discriminant analysis (LDA, Vermeesch 2006a) and the application of classification trees (Vermeesch 2006b) to statistically determine the tectonic affinity of oceanic basalts. LDA has also been implemented by other authors (Agrawal et al. 2004; Agrawal et al. 2008; Verma et al. 2013) for the solution of similar problems.

More recently, Li et al. (2015) further highlighted the inaccuracy of binary and ternary diagrams as discriminating tools in assigning tectonic settings starting from geochemical analyses of igneous rocks. As an alternative, Li et al. (2015) suggested the use of primordial mantle normalized diagrams as qualitative discriminating tools for the different tectonic environments.

In this study, we introduce ML basics and its potential in petrological and mineralogical applications. A ML algorithm is then applied in the attempt to discriminate magmas from different tectonic settings using geochemical data (major elements, trace elements and isotopic data) as input parameters. Although this is a long-standing, highly debated and controversial geological problem, a renewed interest in its solution has developed in recent times among petrologists (see e.g. Li et al. [2015] and references therein). The aim of our work is to cut the Gordian knot representing the complexity associated with the identification of tectonic environments starting from the geochemical composition of magmas.

To provide a robust and quantitative tool we extract a large number of samples from open-access and comprehensive global petrological databases like PetDB (http://www.earthchem.org/petdb) and GEOROC (http://georoc.mpch-mainz.gwdg.de/georoc/). We also test different approaches involving major elements, trace elements and isotopes with the aim of defining the best strategy for the discrimination among different tectonic settings. Finally, we highlight the drawbacks and limitations of the method, and we discuss our results in light of the published literature.

**Methods**

*Basics of Machine learning and potential applications to mineralogy and petrology*

ML is one of the fastest growing areas of computer science, with far-reaching applications (Jordan and Mitchell 2015). One common feature for these applications is that, due to the complexity of the problems that need to be resolved, a human programmer cannot provide an explicit, fine-detailed univocal solution (Shai and Shai 2014). As a consequence, ML algorithms do not try to solve complex problems using an *a-priori* defined conceptual model but they attempt to unravel complexities in large datasets through a so-called learning process (Bishop 2007; Shai and Shai 2014).

The learning process mainly consists of converting experience into "expertise" or "knowledge" (Shai and Shai 2014). Humans use past experiences to implement their learning processes. As an example, a child starts learning the alphabet by looking at any place where he finds a word or a phrase (e.g. a book, a web page, etc.). Then, typically at school, he is taught the meaning of the letters and how to combine them. As a consequence, the learning process for a child is a combination of experiences and teaching. Likewise, the experience for a learning algorithm is the training data and the output is an expertise, which usually takes the form of another computer program and/or a model that can perform a specific task (Shai and Shai 2014).

The previous example allows us to introduce the two main categories of learning in ML: a) unsupervised and b) supervised learning.

In unsupervised applications (such as the case of a child that starts looking at words and letters), the training dataset consists of a number of input vectors without any corresponding target values. On the contrary, in supervised applications (a child learning at school), the training dataset is labeled, meaning that the algorithm learns through examples (Bishop 2007).

Fig. 1 is a flowchart illustrating the main application fields of ML (classification, clustering, regression and dimensionality reduction) and their potential use to solve representative mineralogical and petrological problems. As reported in Fig. 1, a prerequisite for a successful application of a ML technique is the availability of a suitable number of samples (indicatively more than 50). The goal is to define the right ML field (i.e. classification, clustering, regression or dimensionality reduction) in which a problem can be treated (Fig. 1). This involves a series of decisions regarding the nature of problem.

If the problem involves categories, the first step is to select between labeled and un-labeled data. In the case of labeled data, the learning process is supervised and we are dealing with a "classification" problem (Kotsiantis 2007). Potential examples of classification problems in petrology and mineralogy are the classification of crystals in complex systems (e.g. Fischer et al. 2006) and, in the case of this study, the petro-tectonic classification using geochemical data (Pearce et al. 1984). In the case of un-labeled data, we are dealing with a "clustering" problem (Jain et al. 1999). The field of clustering is not completely unexplored in petrology (e.g. Le Maitre 1982). As an example, (Le Maitre 1982) discussed the basics of clustering in petrology. Potential applications of clustering problems in mineralogy and petrology are the discovery of hidden petrological structures in geochemical data or the quantitative analysis of crystal textures (Lach-hab et al. 2010).

If the problem does not involve categories, the next step is to define whether a quantity must be predicted. If the answer is yes, we are in the field of "regression" (Smola and Schölkopf 2004). A potential petrological application of ML regression is the fitting of empirical data from experimental petrology when the mathematical formulation of the problem is not known *a priori*. Example applications are geo-thermometric and -barometric studies or the modeling of evolutionary processes in igneous petrology. Finally, if the aim of the problem does not deal with the prediction of a quantity, we are in the field of "dimensionality reduction" (Lee and Verleysen 2009). This field is particularly useful, for example, in the context of visualization of high-dimensional petrological and mineralogical data. Example applications in petrology are the concepts of composition and reaction spaces introduced by Thompson to solve metamorphic problems (Thompson 1982a; Thompson 1982b).

*Support Vector Machines (SVM)*

Before presenting our analyses, we report some basic information about Support Vector Machines. The readers interested in the details of the SVM theory and numerical methods can find full methodological descriptions in Cortes and Vapnik (1995). A summary of the mathematics behind SVMs is also given in Appendix A.

Support Vector Machines (SVMs) are a set of ML algorithms that are particularly useful in the context of classification (Cortes and Vapnik 1995). During the learning phase, sets of known and categorized training examples are analyzed by the SVM algorithm. The SVM then elaborates a model and assigns unknown samples to different categories (Cortes and Vapnik 1995).

The main strength points of SVMs are: 1) SVMs are effective in high dimensional spaces; 2) SVMs can model complex, real-world problems; 3) SVMs perform well on datasets with many attributes, despite the possible low number of cases which might be available to train the model (Cortes and Vapnik 1995; Yu et al. 2005). All these features make SVMs potentially very useful in the resolution of high-dimensional petrological and geochemical problems of classification.

SVMs numerically implement the following idea: inputs are mapped to a very high-dimension feature space where a decision surface is then constructed (Cortes and Vapnik 1995). In the simplest implementation, the decision surface is linear. It consists of a hyper-plane or set of hyper-planes, which can be used for classification, regression or other tasks. The simplest way to separate two groups of data is by using a straight line (2 dimensions; two chemical elements in our case, and as in the case of binary classification diagrams), a flat plane (3 dimensions, three chemical elements) or an N-dimensional hyper-plane (i.e. N chemical elements). However, certain problems require a non-linear trend to separate the groups more efficiently. SVMs handle these occurrences using non-linear kernel functions (see Appendix A for further details).

SVMs have been originally developed for binary classification problems, where the algorithm learns and performs the classification of unknowns between two classes. However, most of the SVMs applications deal with problems which have a larger number of classes. These applications, such as the case of the discrimination of magmas from different tectonic environments using geochemical data, are defined as multiclass classification problems. In multiclass problems, two main strategies can be implemented: 1) One Vs One (OVO) and 2) One Vs Rest (OVR). In the OVO approach each population is compared with each other population, separately. The OVR approach compares each population with all the other populations simultaneously. Further details about the OVO and OVR approaches are detailed in Appendix A.

*Support Vector Machines Implementation*

To evaluate the use of SVMs in the context of tectonic classification using geochemical data, we used the scikit-learn (Pedregosa et al. 2011) implementation of SVM. Scikit-learn is a Python module integrating a wide range of state-of-the-art machine learning algorithms for medium-scale supervised and unsupervised problems. This package focuses on bringing machine learning to non-specialists using a general-purpose high-level language (Pedregosa et al. 2011). We selected to use the scikit-learn package as this will allow other users to easily replicate our results. In addition, it represents a powerful framework for the solution of petrological and mineralogical problems in fields of clustering, regression, dimensionality reduction and, with regards to the presented case study, classification (Fig. 1). To evaluate the best strategy for the learning stage of the proposed case study, we tested both linear and non-linear kernels using both the OVO and OVR approaches. As non-linear kernel, we selected the so-called Radial Basis Function (RBF) which is one of the most widely used and performing non-linear kernels (Scholkopf et al. 1997).

*Input Data*

For comparison we used the same data sources used in the work of Li et al. (2015). In particular, data are retrieved from the two most comprehensive global petrological databases available to date: PetDB (http://www.earthchem.org/petdb) and GEOROC (http://georoc.mpch-mainz.gwdg.de/georoc/). In order to test SVMs as a general classification tool, we did not limit the present study to basalts but we used all volcanic rock samples for which the whole geochemical characterization of major elements ($SiO_2$, $TiO_2$, $Al_2O_3$, $Fe_2O_{3T}$, $CaO$, $MgO$, $Na_2O$, $K_2O$), selected trace elements (Sr, Ba, Rb, Zr, Nb, La, Ce, Nd, Hf, Sm, Gd, Y, Yb, Lu, Ta, Th) and isotopes ($^{206}Pb/^{204}Pb$, $^{207}Pb/^{204}Pb$, $^{208}Pb/^{204}Pb$, $^{87}Sr/^{86}Sr$ and $^{143}Nd/^{144}Nd$) were available.

Compositions characterized by $SiO_2$ content, on volatile free basis, ranging from 40 to 80 wt.% were selected. Samples marked as altered in the databases were not considered. In addition, we consider the following tectonic environments (Fig. 2), in accordance with Frisch et al. (2011): continental arcs (CA), island arcs (IA), intra-oceanic arcs (IOA), back arc basins (BAB), continental floods (CF), mid-ocean ridges (MOR), oceanic plateaus (OP) and ocean islands (OI) (Table 1). The entire analyzed dataset consists of a total of 3095 samples. Table 1 reports the number of samples from each tectonic setting as well as two statistical indicators (geometric mean and standard deviation, Table 1) for some key geochemical parameters typically used in 'conventional' discrimination diagrams. To better visualize the statistical distribution of some key parameters, Fig. 3 report the histogram distributions for $SiO_2$, Total Alkalies, Zr, La and Y for the different tectonic environments considered in our study. Observing both Table 1 and Fig. 3, it emerges that all the reported parameters show large variations (large standard deviations) and significant overlapping areas among the different tectonic settings, in agreement with the results reported by Li et al. (2015).

*Data Standardization*

Standardization of the dataset is a common requirement for many machine-learning estimators. It involves in reporting each individual feature (i.e. element composition) to a standard, normally distributed population (e.g. Gaussian with zero mean and unit variance). The radial basis function used as non-linear kernel in the present study is one of the estimators requiring a standard, normally distributed dataset (Pedregosa et al. 2011). The standardization process is shown in Fig. 4 using Sr as a representative element. Fig. 4A shows the original, non-standardized, Sr distribution. The Sr composition ranges from 0 to c.a 2800 µg/g and the histogram is characterized an asymmetrical shape (Fig. 4A). To standardize the Sr distribution, we first applied the Box–Cox transformation (Box and Cox 1964; Fig. 4B). Then, the obtained distribution was further transformed by removing the mean and scaling to unit variance (e.g. Templ et al. 2008; Fig. 4C).

*Experiments*

In order to evaluate the classification capabilities of SVMs when dealing with geochemical data, several experiments were performed using major elements, followed by trace elements and isotopes, separately. The major elements were then progressively combined with the trace element and isotopic data. Three sets of experiments have been performed. For the first two experimental sets, the entire dataset was split into two groups containing 70% (learning population) and 30% (test population) of samples. Samples were randomly assigned to the learning and test populations. The learning population was then used to train the ML algorithms and the test population was analyzed as unknown. We first evaluated the learning capabilities of the OVO and OVR strategies using a linear kernel. Successively, we evaluated the performances of a non-linear kernel. Finally, we evaluated the resulting best strategy on the whole dataset using a Leave One Out approach (LOO; James et al. 2013). LOO is one of the simplest cross validation methods and consists in learning the system by taking all the samples except one. The sample which is left out is then introduced within the system as unknown. Thus, for *n* samples, we performed *n* different trainings, one for each unknown sample.

*Metrics*

We define the Classification Score (CS) as the ability of a specific SVM algorithm to discriminate among the different tectonic environments as defined by Frisch et al. (2011) and queried in the reference databases (i.e GEOROC and PetDB; Fig. 2). CS is quantitatively defined as the ratio between the number of correctly discriminated samples and the total number of samples of the test population. A CS score equal to 1 means that the algorithm is capable to recognize and classify all the samples belonging to the test population to the relative tectonic environment (Fig. 2). A CS score equal to 0 means that the algorithm is unable to allocate any sample of the test population to the relative tectonic environment.

**Results and Discussion**

As reported above, in the first sets of experiments a comparison was made between classification capabilities of the OVO and OVR strategies using a linear kernel for major elements (Fig. 5A), trace elements (Fig. 5B), isotopes (Fig. 5C) and the combination of major plus trace elements and isotopes (Fig. 5D) as input variables. In all cases reported in Fig. 5, the CS increases as the number of dimensions of the system increase (i.e. as the number of considered geochemical elements increases). Considering only major elements (Fig. 5A), the classification scores are in range from 0.45 (2D) to 0.66 (8D) and 0.47 (2D) to 0.69 (8D) for the OVR and OVO approaches, respectively. Fig. 5B reports the CS obtained considering only trace elements as input parameters, plotted against the dimension of the system. We reiterate that the dimension of the problem is the number of input parameters and therefore, the number of geochemical elements evaluated by the system. Classification scores for trace elements (Fig. 5B) range from 0.43 (2D) to 0.73 (16D) and from 0.43 (2D) to 0.78 (16D) for the OVR and OVO approach, respectively. Classification scores considering only isotopes (Fig. 5C) range from 0.30 (2D) to 0.45 (5D) and from 0.32 (2D) to 0.47 (5D) for the OVR and OVO approach, respectively. Combining major elements, trace elements and, isotopes (Fig. 5D) the classification score increases progressively reaching values up to 0.84 (29D) and 0.88 (29D) for the OVR and OVO methods, respectively. It is notable that the OVO approach consistently shows higher classification scores relative to the OVR method, evidencing that the OVO approach is more efficient and therefore more suitable than OVR in the classification of the considered samples. We note that the results are in agreement with those reported by Hsu and Lin, (2002), who suggest that OVO methods are more suitable for practical uses relative to other methods. This is due to the fact that, although the OVR approach is faster than OVO, it experiences a bias due to the lack of information regarding the boundaries defining each single population. This does not happen with the OVO approach which, in most cases, provides better results.

At the second stage of our analysis, we evaluated whether the use of a non-linear kernel was able to improve the learning capabilities of the system compared to the linear approach. This was achieved by comparing the results obtained using the RBF kernel function with those obtained from the previous set of experiments, where a linear kernel was utilized. Based on the results of the first experimental set reported above, the OVO approach was used for this second set of experiments. Results are reported in Fig. 6. In this case, classification scores range from 0.53 (2D) to 0.79 (8D), 0.46 (2D) to 0.87 (16D), 0.40 (2D) to 0.79 (5D) and 0.53 (2D) to 0.93 (29D) for major elements (Fig. 6A), trace elements (Fig. 6B), isotopes (Fig. 6C) and the combination of major + trace elements + isotopes (Fig. 6D), respectively.

These results are consistently superior to those obtained for the linear kernel (Fig. 5A-C). To aid in the understanding of the result, those obtained using the linear kernel are reported as a reference in Fig. 6. It is clear that the highest classification score (CS=0.93) is obtained by coupling all the major and trace elements with isotopic data (29D).

According to the above results, we elected to perform a third set of experiments investigating the classification capabilities of the most performing configuration (29D + non-linear kernel + OVO method) in the attempt to determine the tectonic setting of each sample within the dataset. For the third set of experiments the LOO approach was employed. As reported above, it involves learning the system by taking all the samples except one. The sample which is left out is then introduced to the system as an unknown. The procedure is repeated for all the samples belonging to the input dataset and the performance of the experiment is evaluated.

Results are reported in Fig. 7 in the form of a "confusion matrix". This graphical representation is a useful method to display information about actual and predicted classifications allowing a straightforward visualization of results (Provost and Kohavi 1998). In a confusion matrix, each column represents the instances in a predicted class, whereas each row represents the instances in an actual class. As a consequence, correct estimations are reported in the cells belonging to the main diagonal of the matrix; the errors are reported in the other cells. Classification scores higher than 84% are obtained for all the studied tectonic settings with the only exception of BAB samples, for which a score of 65% is obtained (Fig. 7). Classification scores for samples belonging to six tectonic settings are better than 89.5%. They are CA, IOA, IA, CF, OI and MOR characterized by classification scores of 97.3%, 92.0%, 89.5%, 95.0%, 99.2% and 92.4%, respectively. OP shows a lower classification score, but remains above 84%. It is interesting to note that these results are superior to those reported by (Snow 2006) for IA, OI and MOR environments of about 83%, 75% and 76%, respectively.

The case of BAB is intriguing. It is characterized by the lowest classification score (65%) among the considered tectonic settings. This means that about 35% of the samples belonging to the BAB tectonic setting have been misclassified. Among them, significant portions were classified as IOA (27.6%) and MOR (4.1%). Only a few samples were attributed to CA (1.6%), IA (0.8%), CF (0.8%), OP (<0.5%), and OI (<0.5%).

Although the percentage of success in classifying BAB samples is not as large as for the other tectonic environments, noteworthy is the fact that, in the case of a statistically representative population, BAB represents the modal value derived from the classification. We emphasize, however, that the ML system must be used for the evaluation of a statistically representative set of unknown samples and the use of single samples should be avoided. In fact, if one introduces a single BAB sample in the ML system, there is a 65% of probability that it is correctly classified and 35% that it is erroneously classified. Conversely, for example, if one uses a population of 100 BAB samples, a reasonable result might be: 65 BAB, 28 IOA, 4 MOR, 2 CA and one between IA or CF. This points to BAB as the modal value and hence to the most probable petro-tectonic association.

Noteworthy is the fact that, despite the lower percentage of success of BAB classification relative to the other tectonic environments, this represents a step forward compared to previous studies (Saccani 2015; Li et al. 2015) reporting the inability of current methods to discriminate BAB samples. The fact that the main sources of uncertainties are the spreading centers and arch environment is, however, not surprising. This is directly related to the petrological processes governing the genesis and evolution of BAB rocks. In particular, several authors (Taylor and Martinez 2003; Pearce and Stern 2006) highlighted the transitional nature of a large portion of BAB magmatic compositions lying between MOR and arc setting compositions.

The above discussion highlights the great potentials of ML methods in classifying and discriminating the tectonic setting of igneous rocks on the basis of their geochemical and isotopic data. However, although promising, these methods suffer from some limitations that need to be highlighted. In particular, the proposed ML system, as with any other data elaboration technique, must not be considered as a "magic box" where input petrological and geochemical data are transformed into a classification graph or datasheet to be immediately utilized for scientific interpretations. The method needs to be integrated with other techniques such as fieldwork, petrographic observations, classic geochemical studies, geophysical data, etc. to obtain a clear picture of the geologic framework from which samples are collected. Neglecting to do so might compromise a correct interpretation of the ML output exposing the user to some risks. One of the most important is to lose focus about the petrological processes that acted to generate the compositional variability of the samples used in the analyses, irrespective of their correct classification or not. As an example, it might be possible to discover "anomalous" trends of samples using conventional petrological diagrams, such as binary discrimination diagrams. This would allow these trends to be modeled and to advance hypotheses to explain their petrological nature. This task cannot be undertaken using the ML classification scheme proposed in our work. This does not mean, however, that ML techniques cannot be used to model petrological processes. As shown in Fig. 1, the qualitative and quantitative modeling of petrological processes can be achieved by ML techniques in the fields of "dimensionality reduction" and "regression", respectively.

A further limitation of the proposed classification system that must be highlighted is the inability in defining new classification groups that are different from those which are already defined. As an example, samples affected by secondary processes that altered their original geochemical composition will be necessarily classified in one of the available tectonic settings, possibly producing biases and misclassifications. In this case, preliminary petrographic inspections and geochemical investigations can be decisive in excluding those samples from the ML process. It is therefore important to combine classic petrological and geochemical approaches, together with the ML techniques, in order to resolve the multidimensional and complex nature of petrological problems.

**Concluding Remarks**

In this contribution we introduced the potentials of Machine Learning in petrology and mineralogy. We explored the applicability of support vector machines in discriminating among different tectonic environments using a large number of dimensions (up to 29D; i.e. 29 parameters, including major, trace elements and isotopes) and the two largest and comprehensive global petrological databases: PetDB and GEOROC.

We demonstrated that: 1) Machine learning has oustanding potentials in petrological and mineralogical studies; 2) support vector machines are robust and useful tools in addressing the complexity underlying the extreme compositional variability often encountered in petrological studies; 3) support vector machines are able to discriminate among different tectonic settings; 4) trace elements alone display a good discriminatory power for the different tectonic settings (CF=0.87); 5) the combination of major elements, trace elements and isotopic data provides much more reliable results (CF=0.93) compared to methods in which these three groups of geochemical parameters are analyzed separately (CF=0.79 for major elements, CF=0.87 for trace elements and CF=0.79 for isotopes).

With regards to the case study, our model is able to discriminate the tectonic setting for studied rock samples with high percentage of success. The only exception is the dataset from back-arc basins that shows lower percentages of success. This is interpreted as a result of the transitional nature of a large proportion of back-arc basin magmatic compositions lying between mid-ocean ridge and arc setting compositions.

The results of this study can have important implications in petrology opening new perspectives for petrologists and geochemists. A potential application is the study of the large amount of data arising from experimental petrology in order to find new relations and uncovered patterns. Another example is the development of more robust geo-thermometers and/or geo-barometers by joining results deriving from different techniques. A further example of application might be the determination of the provenance of crypto-tephra using geochemical data, in order to correlate distal tephra to past volcanic events. Until now, as in the case of tectonic environment determination, geochemical tephro-chronological correlations are mainly performed using binary or ternary diagrams (Tomlinson et al. 2015). The combined use of large databases coupled with machine learning techniques might provide a more robust quantitative approach to better correlate crypto-tephra layers to volcanic sources.

**Acknowledgments**

We thank the editor (Prof. O. Müntener) and two unknown reviewers for valuable comments and suggestions that contributed to increase the quality our manuscript. We also acknowledge Rebecca Astbury for the proofreading of the final version of the manuscript. This project was supported by the ERC Consolidator "CHRONOS" project (grant no. 612776) and by the Microsoft Research Azure Award Program (Maurizio Petrelli: Azure Machine Learning Award).

**Appendix A: mathematical principles of support vector machines**

An extensive introduction to the mathematical principles of support vector machines is reported in Abedi et al. (2012) and Cortes and Vapnik (1995). To introduce the formulation of support vector machines we first discuss a two-class problem.

Consider a training dataset of S dimensional samples (e.g. S chemical elements as input) $x_i$ with $i=1,2,3,…,n$ where $n$ is the number of samples. To each sample, a label $y_i$ is assigned. The label $y_i$ is equal to 1 for the first class and -1 for the second class.

In the case the two classes are linearly separable, then there exists a group of linear separators that satisfy the following equation (Kavzoglu and Colkesen 2009):

$$w.x_i + b \geq +1 \quad for \quad y_1 = +1$$

$$w.x_i + b \leq -1 \quad for \quad y_1 = -1$$

As a consequence, the separating hyper-plane can be formalized as a decision function:

$$f(x) = sgn(wx + b)$$

with *sgn(x)* defined as follow:

$$sgn(x) = \begin{cases} 1 & if \ x > 0 \\ 0 & if \ x = 0 \\ -1 & if \ x < 0 \end{cases}$$

The parameters of *w* and *b* can be obtained by solving the optimization function:

$$minimize \ \tau(w) = \frac{1}{2}\|w\|^2$$

subject to:

$$y_i\big((wx_i) + b\big) \geq 1, \quad i = 1,2,3,\dots,n$$

An example of two-dimensional problem where two different populations can be divided by a linear function is reported in Fig. A1A. However, there are problems where a non-linear trend can separate the different populations more efficiently (Fig. A1B).

In these cases, a projection function $\phi(x)$ can be utilized to map the training data form the original space *x* to a Hilbert space *X*. This means that a non-linear function is learned by a linear learning machine in a high-dimensional feature space while the capacity of the system is controlled by a parameter that does not depend on the dimensionality of the space (Cristianini and Shawe-Taylor 2000). This is called "kernel trick" and means that the kernel function transforms the data into a higher dimensional feature space allowing for performing a linear separation (Cortes and Vapnik 1995).

As reported by Abedi et al., (2012) the training algorithm in the Hilbert space only depend on data in this space through a dot product (i.e. a function with the form $\langle \phi(x_i) \times \phi(x_j) \rangle$). As a consequence, a kernel function $K$ can be formalized as follows:

$$K(x_i, x_j) = \langle \phi(x_i) \times \phi(x_j) \rangle$$

The two-class problem can be also solved as follow (El-Khoribi 2008):

$$maximize \sum_{i=1}^{n} \alpha_i - \frac{1}{2} \sum_{ij=1}^{n} \alpha_i \alpha_j y_i y_j K(x_i, x_j)$$

subject to:

$$\alpha_i \geq 0, \quad i = 1,2,3, \ldots, n \quad and \quad \sum_{i=1}^{n} \alpha_i y_i = 0$$

The decision function can be now rewritten as (Yang et al. 2008):

$$f(x) = sgn\left(\sum_{i=1}^{n} y_i \alpha_i K(x_i, x_j)\right)$$

Many potential functions can be utilized as $K(x_i, x_j)$ (Zuo and Carranza 2011). Among these, the radial basis function (RBF) utilized in our work is defined as follow:

$$K(x_i, x_j) = e^{-\gamma(x_i - x_j)^2}$$

As reported by Cortes and Vapnik (1995), support vector machines were originally developed for the solution of two-class problems, but many of the potential applications are characterized by more than two classes (multi-class problems). In order to solve multiclass problems, the two most popular approaches are the One Vs

One (OVO) and the One Vs Rest (OVR) approach (Fig. A2). In OVO, one SVM classifier is built for all possible pairs of classes (Fig. A2B and A2C) (Knerr et al. 1990; Dorffner et al. 2001). The output from each classifier is obtained in the form of a class label. The class label with the highest frequency is assigned to that point in the data vector (Hsu and Lin 2002). Since the number of SVMs required in this approach is $M(M-1)/2$, it is not suitable for those datasets characterized by a large number of classes (Dorffner et al. 2001).

On the contrary, in OVR one SVM is built for each of the M classes. The SVM for a particular class is constructed using the training examples from that class as positive examples and the training examples of the rest of (M-1) class as negative examples (Fig. A2D).

In other words, in the OVO (One Vs One) approach each population is compared with each other population, separately. In the OVR (One Vs Rest) approach each population is compared with all the other populations mixed together, simultaneously.

**Appendix B: the logic behind classification**

Fig. B1 reports a flowchart showing the steps to be implemented to determine the tectonic environment of igneous. The first step consists in verifying whether the learning process has been already performed. In the case the learning is missing, a new learning process is required. To complete this task, the reference dataset has to be normalized and split into two portions: the learning and test dataset. The role of the learning dataset is to train the system and develop a provisional model. The role of the test dataset is to check the goodness of the provisional model developed using the learning dataset. To complete this task, the samples belonging to the test dataset are evaluated as unknowns using the provisional model. If the validation process is completed successfully, the provisional model is converted to a final model. On the contrary, the whole classification process is aborted and more detailed studies are required.

When the final model is ready, the samples belonging to the unknown population are processed by the system. Results are then cross-validated using conventional techniques such as petrographic inspections, classical geochemical investigations, field observations etc.

Finally, if the provisional results are confirmed, the unknown samples can be safely assigned to a specific tectonic setting. On the contrary, further investigations are needed.

**Figure Captions**

Figure 1: Flowchart showing the main application fields of machine learning (classification, clustering, regression and dimensionality reduction) and their applicability to representative petrological and mineralogical problems. The figure also shows how to select a specific application field in the attempt to solve a petrological or mineralogical problem.

Figure 2: Definition of the considered tectonic environments in accordance with Frisch et al. (2011). As a reference, typical magmatic regions are also indicated.

Figure 3: Histograms for selected key parameters ($SiO_2$, Total Alkalies, Zr, La and Y) for the different tectonic environments: continental arcs (CA), island arcs (IA), intra-oceanic arcs (IOA), back arc basins (BAB), continental floods (CF), mid-ocean ridges (MOR), oceanic plateaus (OP) and ocean islands (OI).

Figure 4: Illustration of the normalization procedure using Sr as a representative element: A) original univariate histogram distributions; B) histogram distributions after the application of the Box-Cox transformation; C) histogram distributions after the subtraction of the mean and the transformation to get a variance equal to 1.

Figure 5: Classification scores for the linear kernel plotted against the number of dimensions (D; i.e. number of chemical elements or isotopes) for the OVO and OVR approaches; A) results for major elements; B) results for trace elements; C) results for isotopes; D) result for the combination of major elements + trace elements + isotopes.

Figure 6: Classification scores for the non-linear kernel plotted against the number of dimensions (D; i.e. number of chemical elements or isotopes) using the OVO approach; A) results for major elements; B) results for trace elements; C) results for isotopes; D) result for the combination of major elements + trace elements + isotopes. Results for the linear kernel are also reported for comparison.

Figure 7: Confusion matrix of the LOO cross calibration using major elements + trace elements + isotopes (29D), the non-linear kernel and the OVO approach. See text for details.

Figure A1: Simplified examples of 2D populations that can be separated by a linear (A) and non-linear (B) functions.

Figure A2: Exemplification of the OVR and OVO approaches in 2D for the discrimination among the three populations reported in A. In OVO (B and C) each population is compared with each of the population separately (M-1) times; in OVR (D) each population is compared against all the other populations mixed together.

Figure B1: Flowchart showing the steps to be performed in the attempt of determining the tectonic setting of unknown samples.

**Table Captions**

Table 1: Synoptic table for the considered tectonic environments. The number of samples utilized in our study for each environment, the geometric mean and the standard deviation of some key parameters ($SiO_2$, Total Alkalies, Mg#, Zr, Nb, La, Y) are shown.

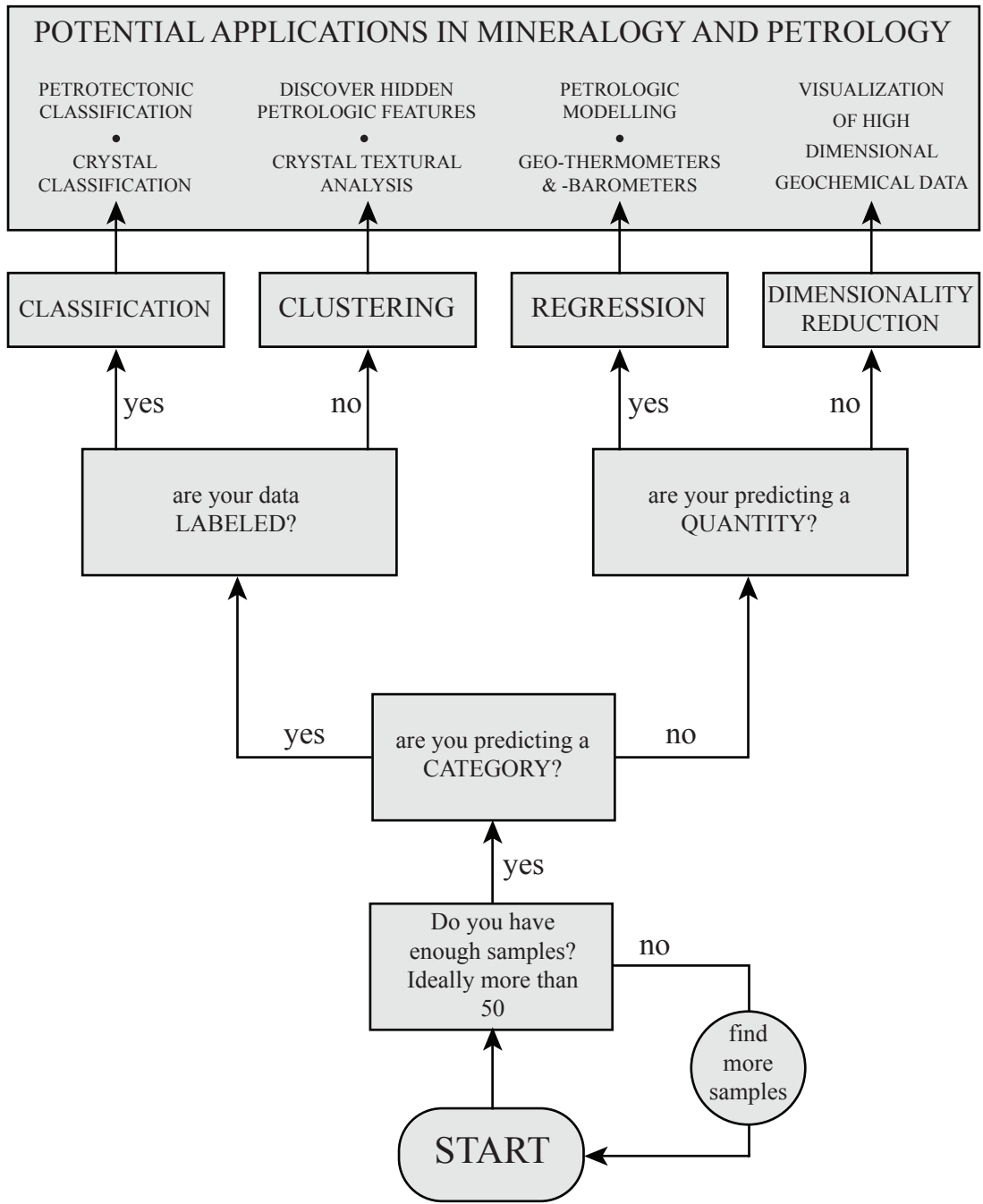

Figure 1

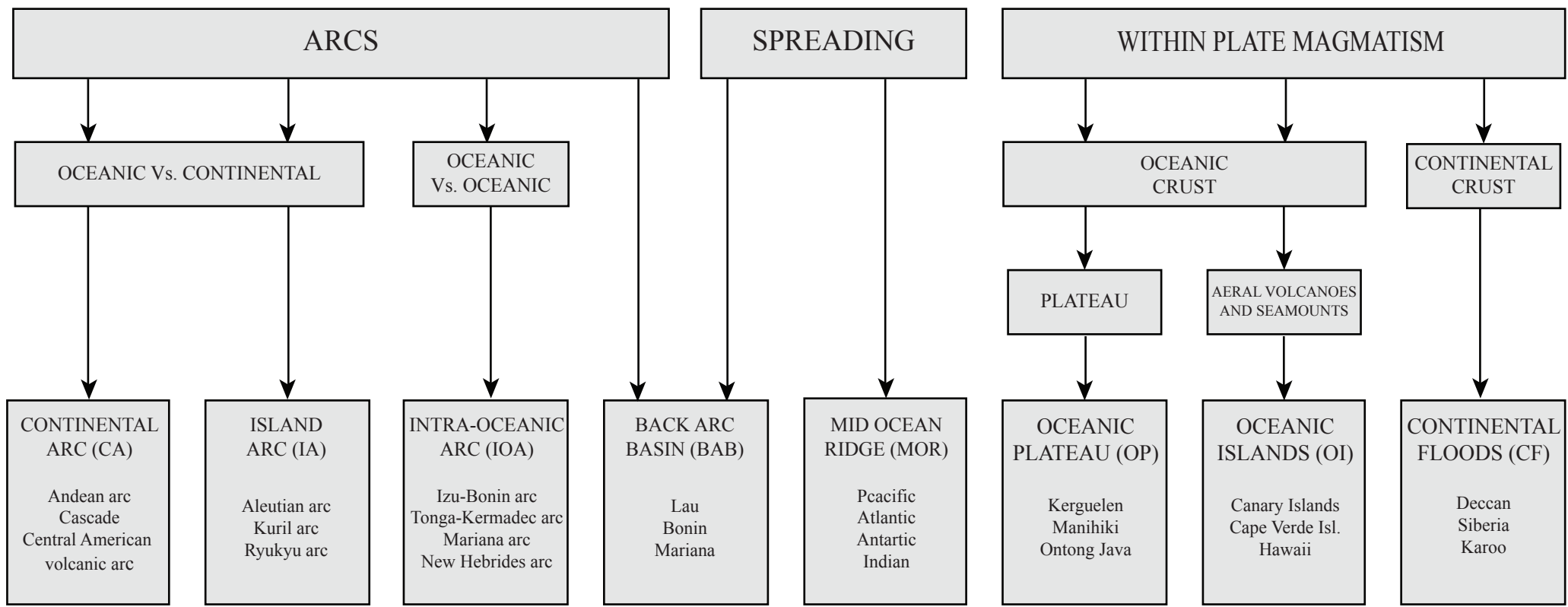

Figure 2

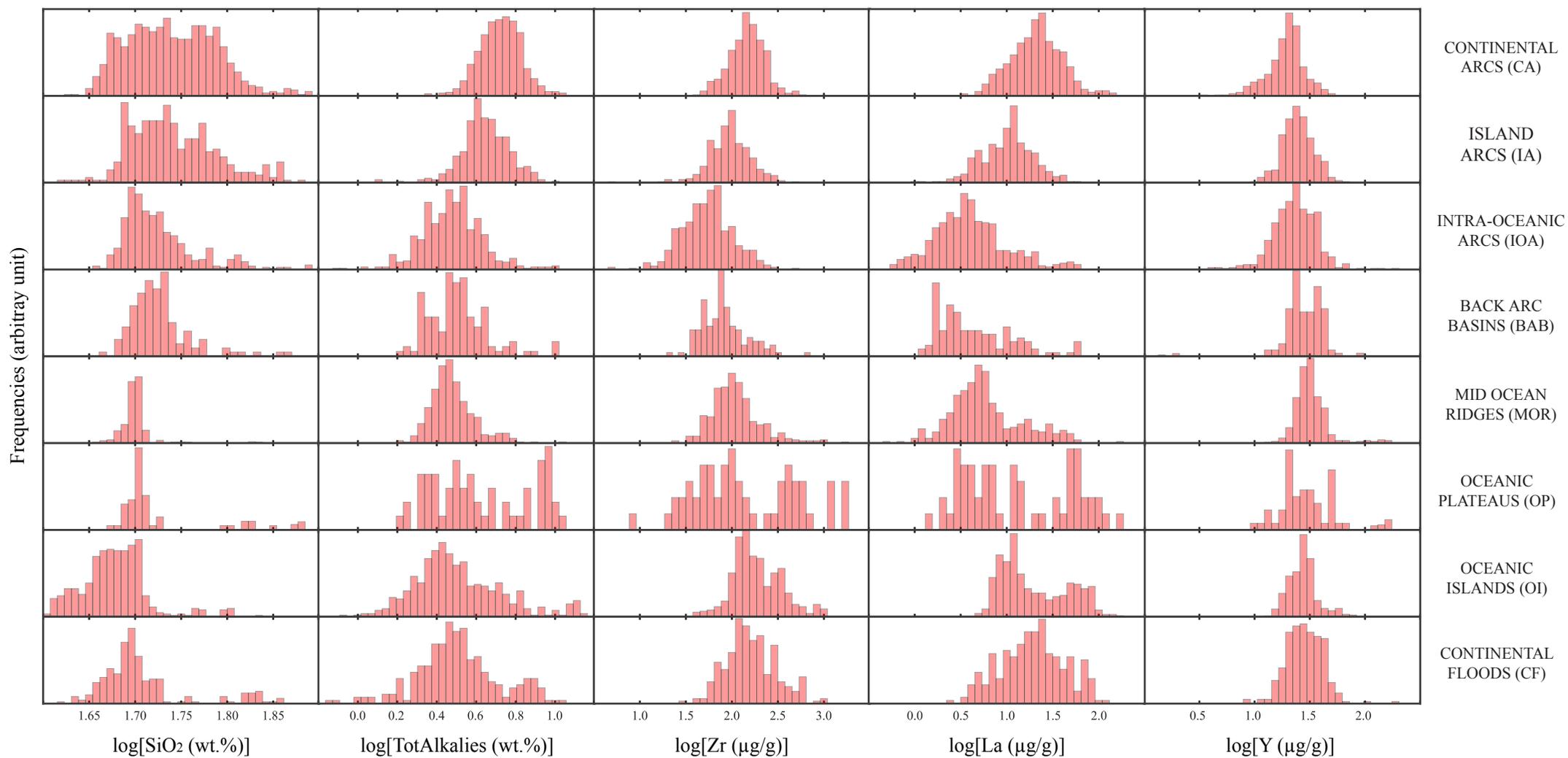

Figure 3

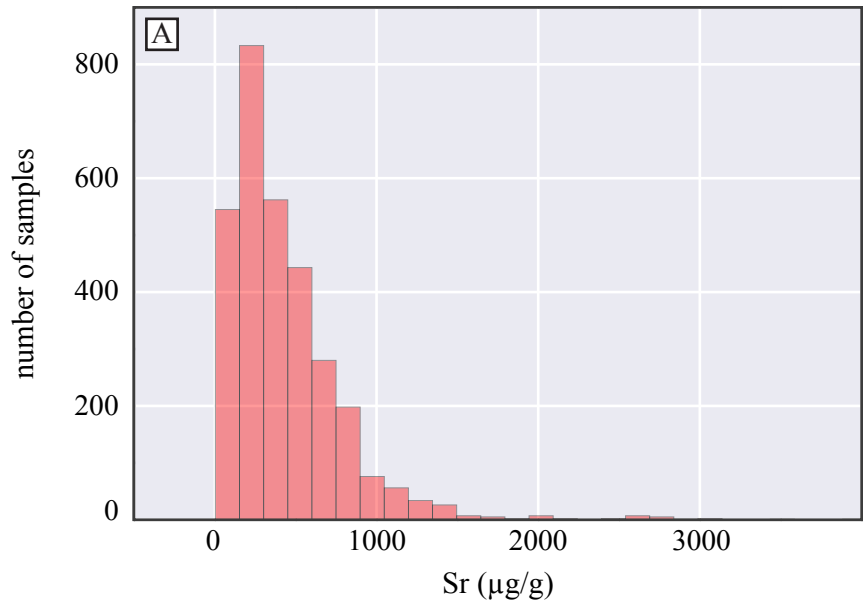

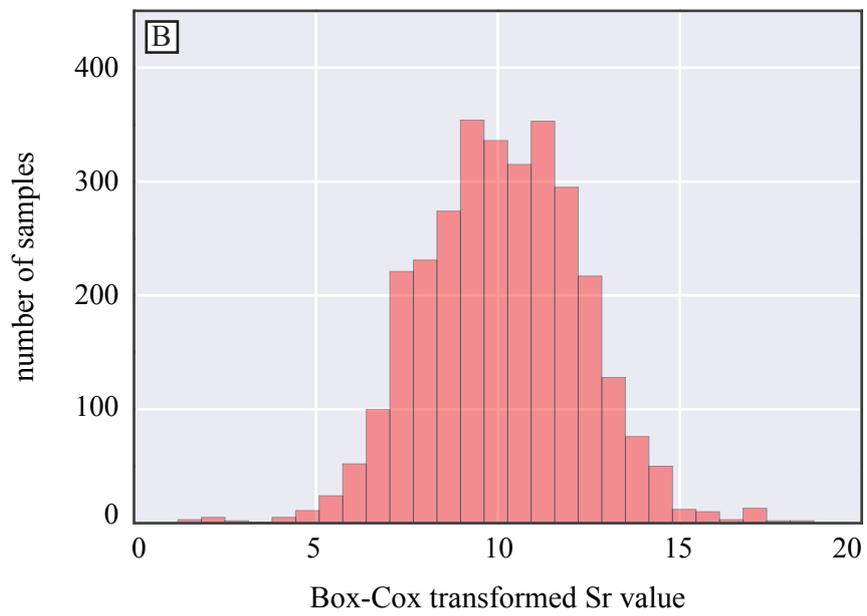

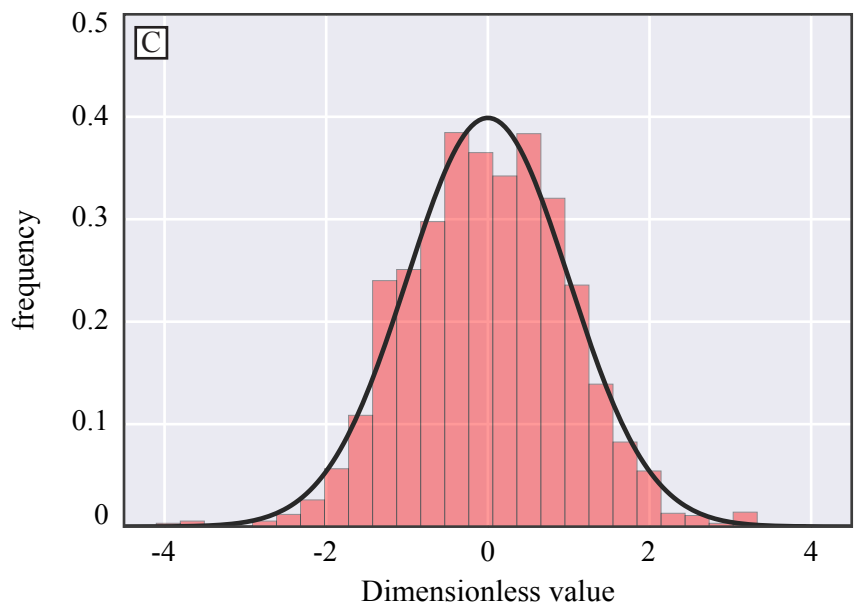

Figure 4

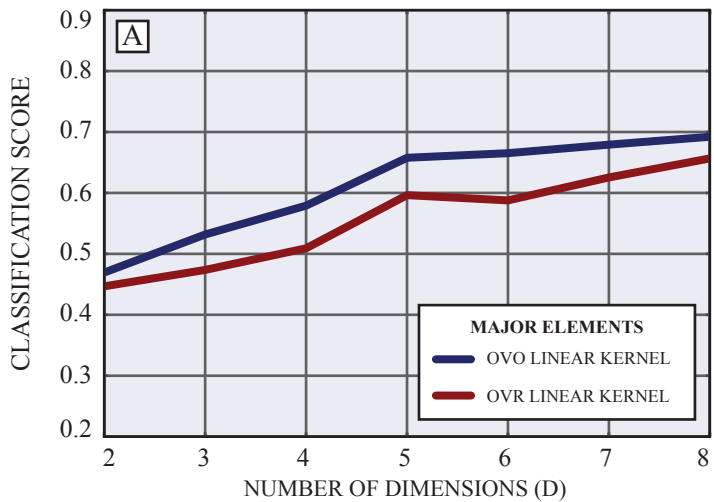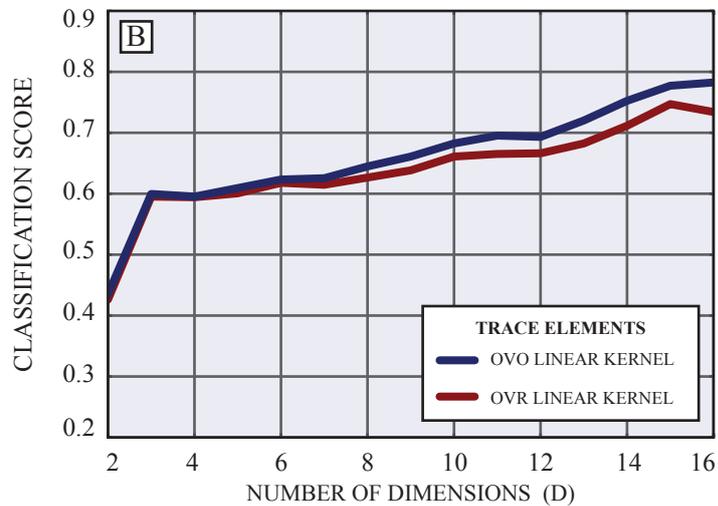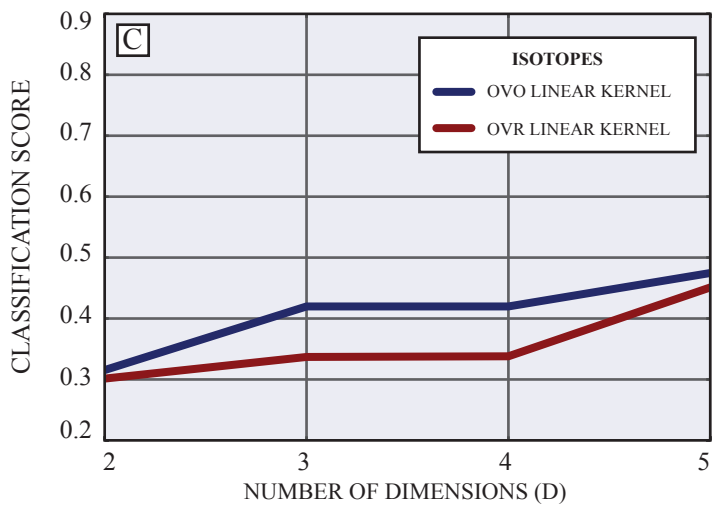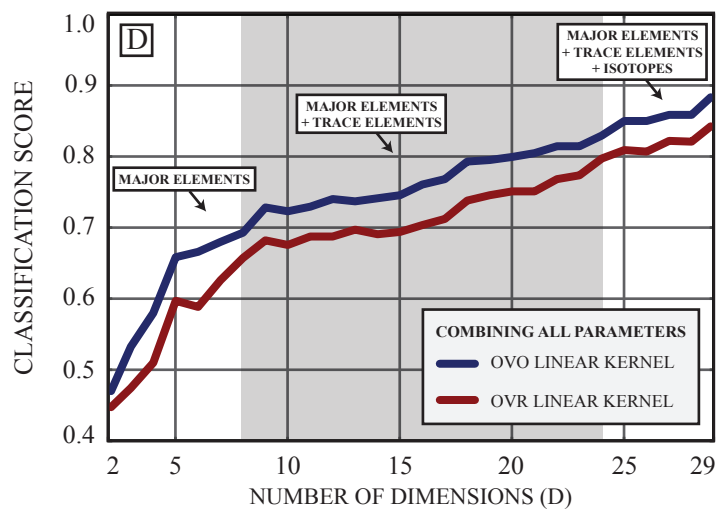

Figure 5

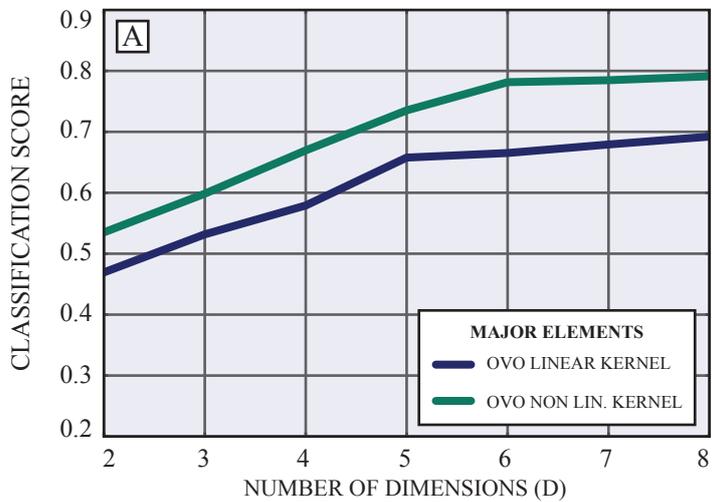
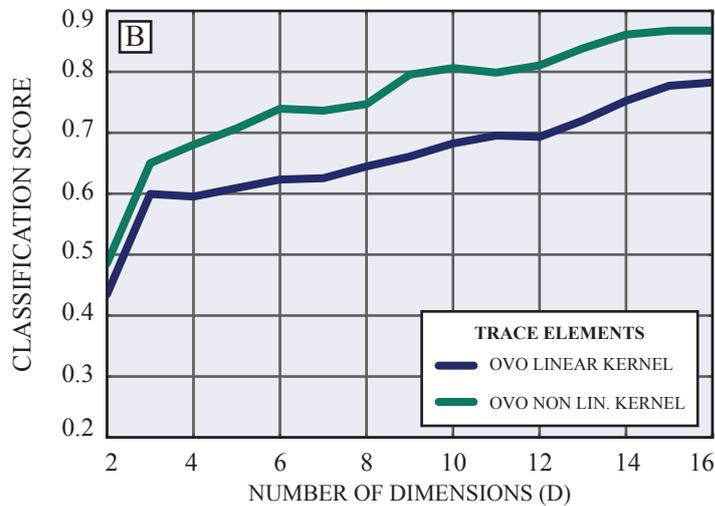
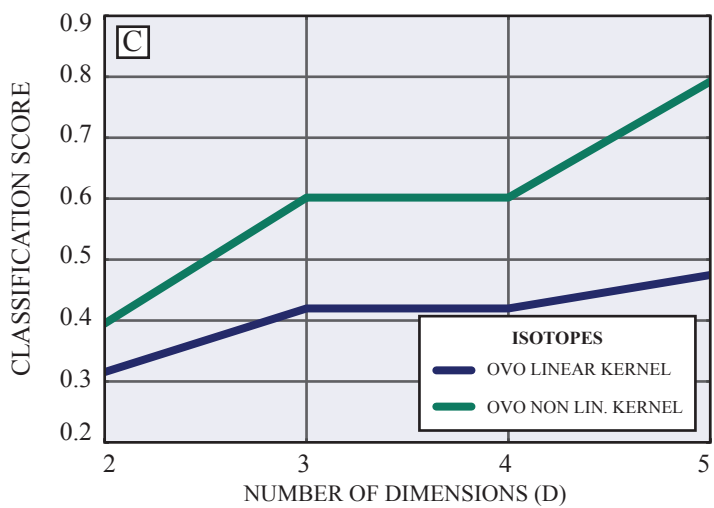
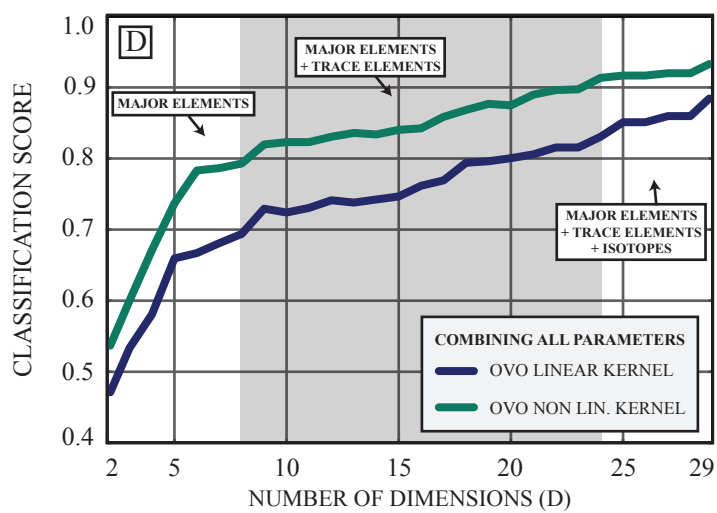

Figure 6

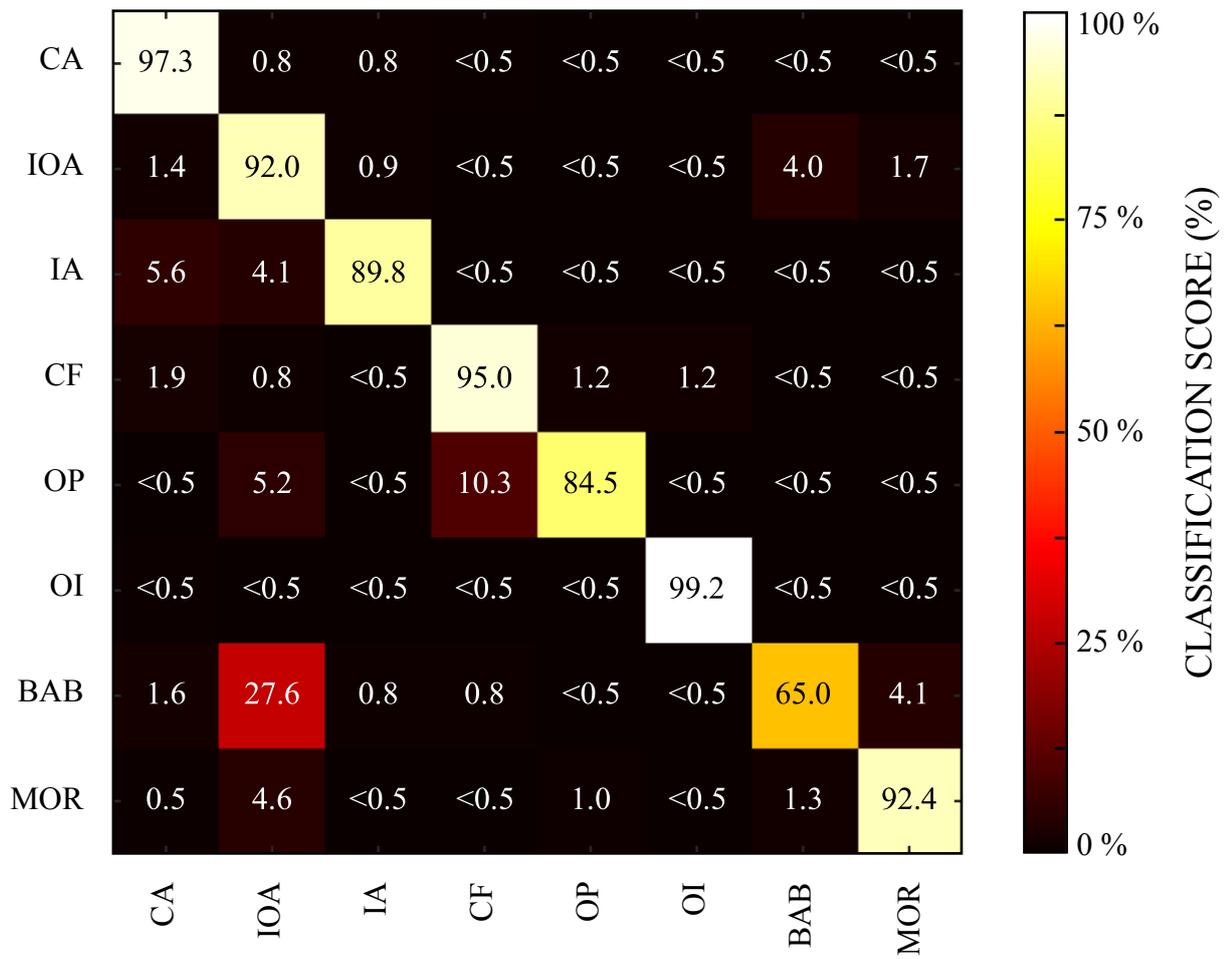

Figure 7

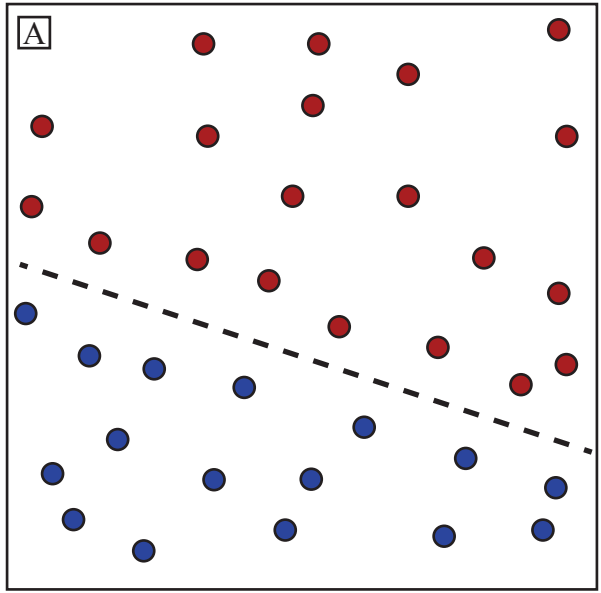

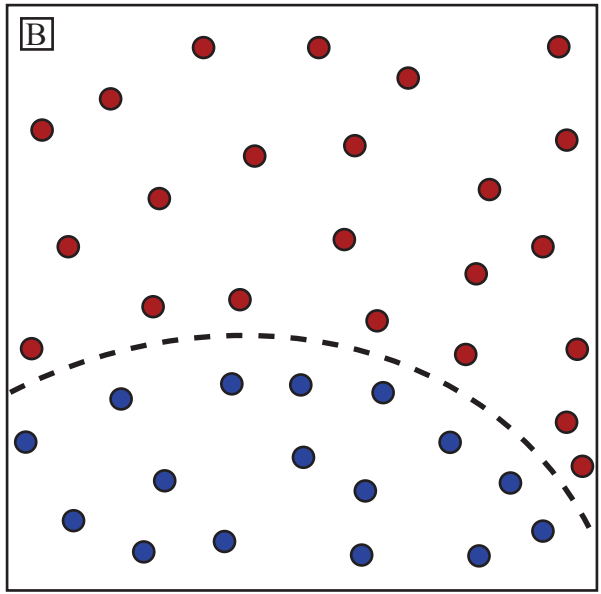

Figure A1

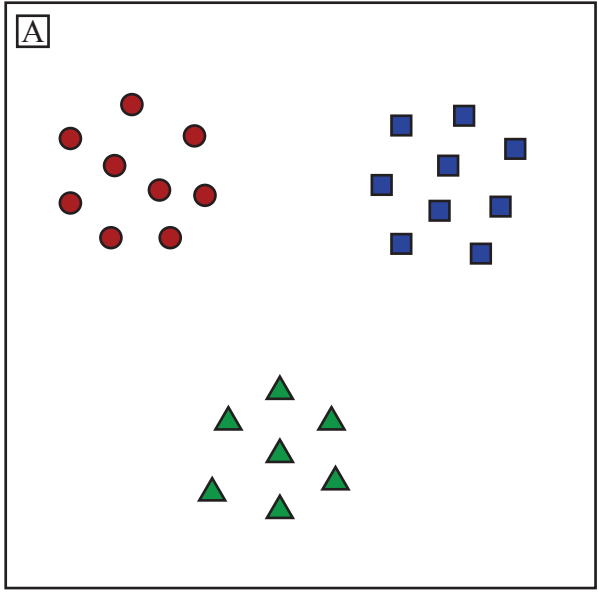
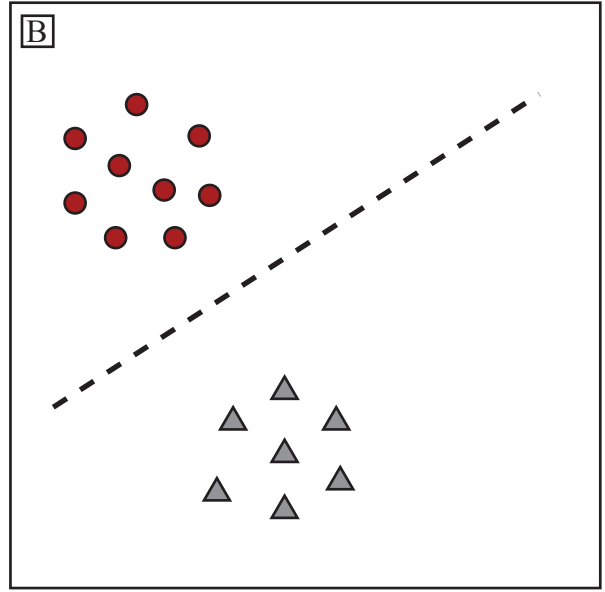
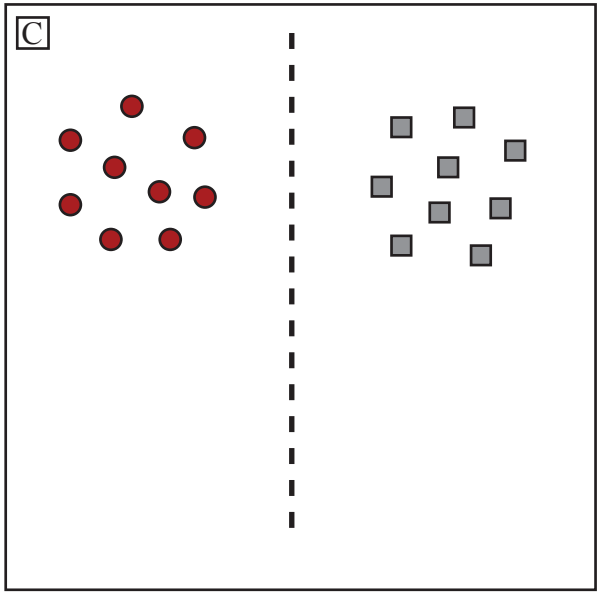
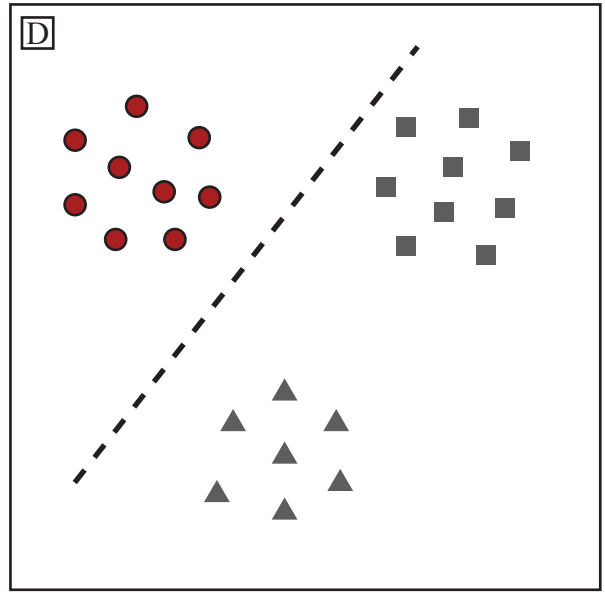

Figure A2

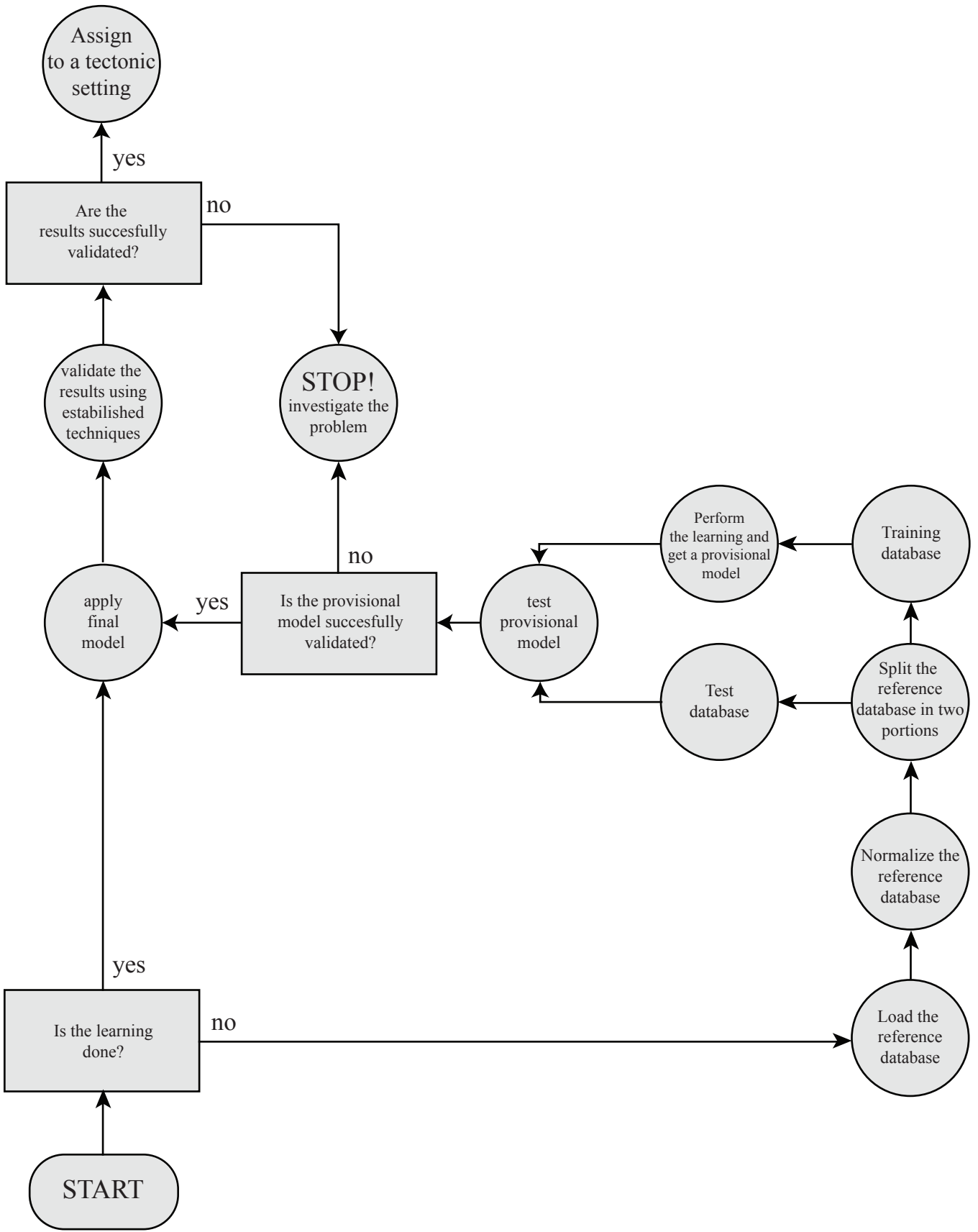

Figure B1

| Tectonic Setting | n. of samples | SiO$_2$ (wt.%) | 1σ | Total Alkalies (wt.%) | 1σ | Mg # | 1σ | Zr (µg/g) | 1σ | Nb (µg/g) | 1σ | La (µg/g) | 1σ | Y (µg/g) | 1σ | Compositions |
|---|---|---|---|---|---|---|---|---|---|---|---|---|---|---|---|---|
| Continental Arc (CA) | 840 | 55 | 6 | 5 | 1 | 51 | 9 | 146 | 76 | 9 | 12 | 21 | 20 | 19 | 8 | Basaltic to dacitic |
| Intra-oceanic Arc (IOA) | 654 | 54 | 5 | 3 | 1 | 47 | 12 | 56 | 45 | 1 | 4 | 4 | 8 | 23 | 13 | Mostly basaltic and andesitic, subordinate dacites |
| Island Arc (IA) | 266 | 55 | 7 | 5 | 1 | 46 | 10 | 105 | 63 | 3 | 11 | 12 | 8 | 26 | 9 | Mostly basaltic and andesitic |
| Back Arc Basins (BAB) | 123 | 53 | 4 | 3 | 1 | 47 | 13 | 82 | 80 | 2 | 4 | 4 | 12 | 26 | 11 | Basaltic, subordinate andesites and dacites |
| Continental Floods (CF) | 258 | 51 | 6 | 3 | 2 | 51 | 15 | 162 | 152 | 13 | 19 | 19 | 21 | 28 | 15 | Basaltic, subordinate andesites |
| Mid-Ocean Ridge (MOR) | 394 | 50 | 3 | 3 | 1 | 55 | 10 | 110 | 147 | 6 | 22 | 6 | 13 | 33 | 20 | Mostly basaltic |
| Oceanic Plateau (OP) | 58 | 54 | 8 | 4 | 3 | 43 | 21 | 143 | 440 | 13 | 44 | 13 | 37 | 31 | 34 | Basaltic, subordinate andesites and dacites |
| Ocean Island (OI) | 502 | 48 | 4 | 3 | 3 | 57 | 13 | 188 | 175 | 23 | 38 | 20 | 27 | 27 | 10 | Basaltic, subordinate andesites |

Table 1